\documentclass[pdflatex,sn-mathphys]{sn-jnl}
\usepackage[utf8]{inputenc}
\usepackage{bbm}
\usepackage{amsmath}
\jyear{2022}
\newtheorem{prop}{Proposition}
\newtheorem{cor}{Corollary}
\newtheorem{ex}{Example}

\title[Power Transformations of Relative Count Data]{Power Transformations of Relative Count Data as a Shrinkage Problem}

\begin{document}

\author*{\fnm{Ionas} \sur{Erb}}\email{ionas.erb@crg.eu}

\affil{\orgdiv{The Barcelona Institute of Science and Technology}, \orgname{Centre for Genomic Regulation (CRG)}, \orgaddress{\street{C/ Dr Aiguader, 88}, \city{Barcelona}, \postcode{08003}, \country{Spain}}}


\abstract{Here we show an application of our recently proposed information-geometric approach to compositional data analysis (CoDA). This application regards relative count data, which are, e.g., obtained from sequencing experiments. First we review in some detail a variety of necessary concepts ranging from basic count distributions and their information-geometric description over the link between Bayesian statistics and shrinkage to the use of power transformations in CoDA. We then show that \emph{powering}, i.e., the equivalent to scalar multiplication on the simplex, can be understood as a shrinkage problem on the tangent space of the simplex.  In information-geometric terms, traditional shrinkage corresponds to an optimization along a mixture (or $m$-) geodesic, while powering (or, as we call it, \emph{exponential} shrinkage) can be optimized along an exponential (or $e$-) geodesic. While the $m$-geodesic corresponds to the posterior mean of the multinomial counts using a conjugate prior, the $e$-geodesic corresponds to an alternative parametrization of the posterior where prior and data contributions are weighted by geometric rather than arithmetic means. To optimize the exponential shrinkage parameter, we use mean-squared error as a cost function on the tangent space. This is just the expected squared Aitchison distance from the true parameter. We derive an analytic solution for its minimum based on the delta method and test it via simulations. We also discuss exponential shrinkage as an alternative to zero imputation for dimension reduction and data normalization.}




\maketitle

\section{Introduction}

\begin{figure}[h]%
\centering
\includegraphics[width=0.6\textwidth]{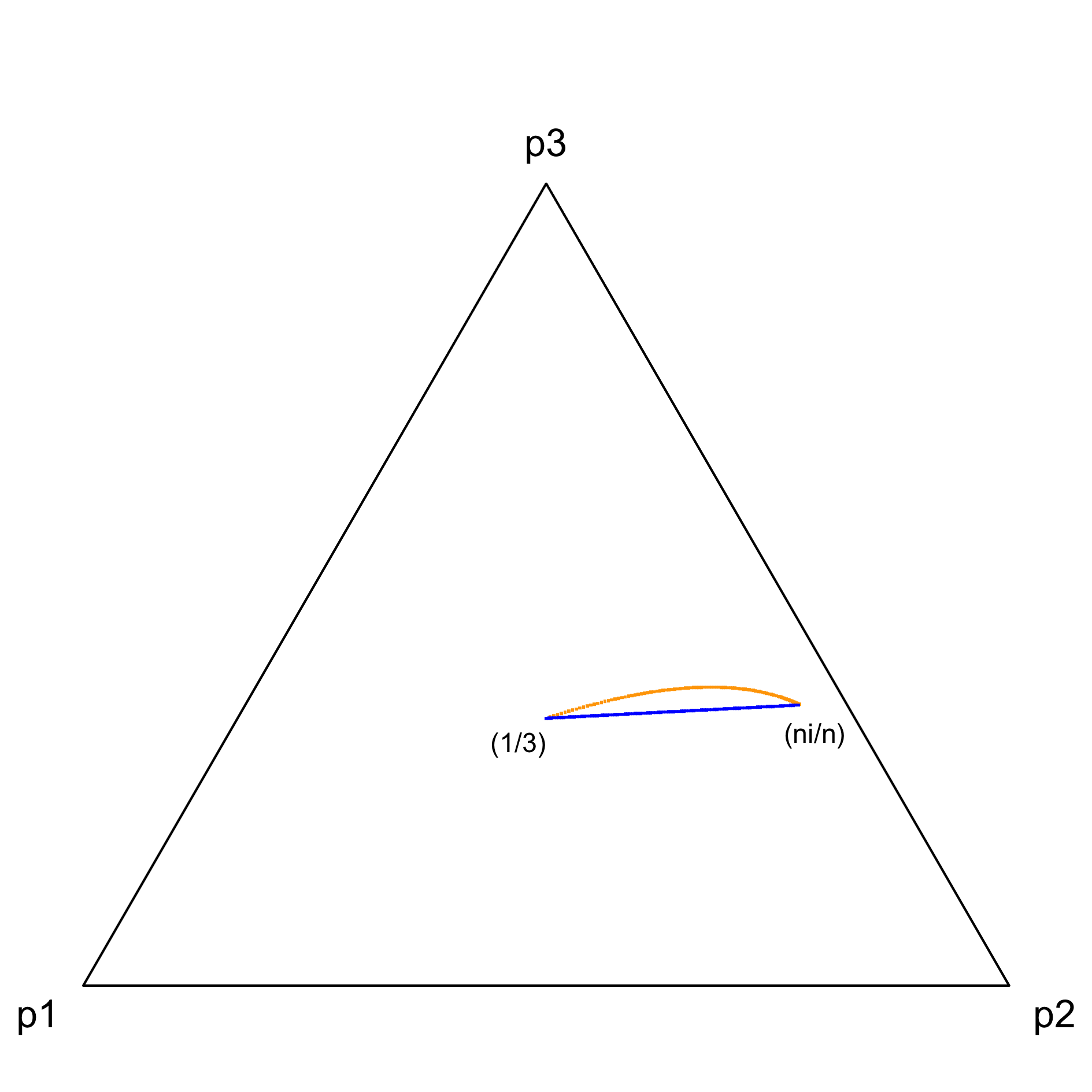}
\caption{Exponential (curved orange line) and mixture (blue straight line) geodesics between the equidistribution $(1/3,1/3,1/3)$ and an observed point $(n_1/n,n_2/n,n_3/n)$ in the 3-part simplex.}\label{geodesics}
\end{figure}

Counting discrete events seems one of the simplest ways of collecting data, but compositional bias when directly comparing such counts in varying contexts can lead intuition astray. Often, the lack of a common scale in samples taken from different environments or experimental conditions makes direct comparisons between counts meaningless. We need to gauge by internal references before we can make external comparisons. Compositional data analysis (CoDA, e.g. \cite{Greenacre2021}) uses scale-free methods on data occurring in form of percentages, and its log-ratio methodology \cite{AitchisonBook} has been applied to relative counts as well. While the sample spaces \cite{sampleSpace} of both data types are certainly not the same, the underlying problematic is identical: direct comparisons across samples can have paradoxical effects due to the lack of a common scale \cite{specialIssue}. We have recently proposed to make  use of information geometry \cite{Amari} to analyse compositional data \cite{ErbAy}. The information-geometric approach is even more natural for relative count data, and simple count distributions like the categorical or multinomial have served as examples to illustrate basic concepts in information geometry. Here we aim to demonstrate the usefulness of information-geometric concepts for the analysis of count data that are compositional in a well-defined sense.\\
Let us quickly sketch the main idea of this contribution. Consider a vector of counts $(n_i)_{i=1}^D$ that were produced by some process with unknown independent count probabilities $q_i$. It is well known that the empirical estimator for such multinomial probabilities 
\begin{equation}
    \hat{q_i}=\frac{n_i}{\sum_{k=1}^Dn_k}
\end{equation}
(although it is the one that maximizes the likelihood of the data) can be much improved upon when the denominator is not large compared with $D$. In this case, a better alternative is the convex combination
\begin{equation}
    \hat{q_i}^\mathrm{sh}=\lambda\frac{1}{D}+(1-\lambda)\hat{q_i}\label{shrink}
\end{equation}
of the estimator with the equidistribution,  for an optimized value of the parameter $0\le\lambda\le1$. This is an example of what is known as  {\it shrinkage} of $\hat{q}_i$ toward the target $1/D$. The reason why this works can be understood from a Bayesian perspective. The shrinkage estimator (\ref{shrink}), instead of maximizing the likelihood of the data, maximizes the posterior probability of a suitable parameter of the multinomial (assuming a simple conjugate prior). Optimization of $\lambda$ corresponds to adjusting the weight that the prior will have compared with the weight that will be assumed for the data. But why is {$\hat{q}_i^\mathrm{sh}$} a good approximation of $q_i$? It turns out that maximizing the posterior probability corresponds to minimizing the divergence of $\hat{q}_i^\mathrm{sh}$ from $q_i$.\\
As the parameters (and estimators) we are dealing with are probabilities themselves, they can be understood as points in a finite simplex (which happens to be the CoDA sample space). From an information-geometric point of view, the shrinkage estimator is optimized along the {\it mixture geodesic} (or $m$-geodesic) between the equidistribution and the observed point $(\hat{q}_i)_{i=1}^D$ (see the blue line in Figure \ref{geodesics}). Geodesics provide intuition, e.g., a generalized Pythagorean theorem makes use of them. Unlike in Euclidean geometry, however, we need \emph{two} types of geodesics for Pythagoras to work. The natural counterparts to $m$-geodesics are the {\it exponential geodesics} (or $e$-geodesics). These are convex combinations of points in exponential coordinates, which are dual to the mixture coordinates (via the Legendre duality that underlies information geometry). Let us now consider the $e$-geodesic between the two points in question (see the orange curve in Figure \ref{geodesics}).
It turns out that the $e$-geodesic corresponds to an alternative parametrization of the posterior probability, where the prior and likelihood contribute via weighted geometric means. A point on the $e$-geodesic is just another estimator of the posterior mean that uses this alternative parametrization. When back-transforming exponential coordinates to the original parameter, this geodesic can be written as
\begin{equation}
    \hat{q}_i^\mathrm{es}=\frac{\hat{q}_i^\beta}{\sum_{k=1}^D\hat{q}_k^\beta},
\end{equation}
with $0\le\beta\le1$. This kind of exponential scaling is well known in statistical physics, where $\beta$ is the inverse temperature. It is also used when Box-Cox transforming data to reduce skew or to replace logarithms by approximate expressions when zeros are involved. In the CoDA context, $\beta$ can be used to mediate between $\chi$-squared distance and Aitchison distance and thus makes a connection between log-ratio analysis and Correspondence Analysis (CA) \cite{Greenacre2010}. The latter can handle zeros while the former needs to impute them.\\
For finding the optimal value of the shrinkage parameter $\lambda$, a simple analytic solution for minimization of the mean squared error (MSE) with respect to the true parameter can be found \cite{LedoitWolf, HausserStrimmer}. To use the same strategy for the $\beta$-parameter of the $e$-geodesic, we propose to use an MSE on the tangent space. This is just expected Aitchison distance between the estimator and the true parameter. We derive an analytic solution that approximates an optimal $\beta$ based on the Delta method (i.e., via Taylor expansion). This is computationally inexpensive and can, e.g., be used as a data preprocessing for dimension reduction techniques like CA. Simulations show that this approach holds promise for data with many essential zeros. We discuss the exponential shrinkage estimator as an additional tool that avoids the pseudocounts of current procedures in contexts where zero imputation may be inappropriate. On a theoretical level, this contribution aims to unify power transformations with shrinkage under the same conceptual framework.\\
{Section \ref{review} presents essentially review material, with the first two paragraphs dedicated to some very general statistical motivation. We then introduce the information geometric formulation of the multinomial likelihood and posterior and make some methodological excursions of a  more technical nature in paragraphs \ref{generalEst} and \ref{shrinkage}. In these paragraphs, we reformulate known minimizations of relative entropy and of expected quadratic loss in form of propositions that will serve us in the subsequent application. Section \ref{application} is then dedicated to the application of the material presented. It includes the definition of an alternative shrinkage estimator and its optimisation along the exponential geodesic as well as a benchmark of it using simulations. All the proofs and some of the more lengthy algebraic derivations are deferred to the Appendix.}

\section{Preliminaries}\label{review}

\subsection{Sequencing data are relative}
Let us first discuss the practical relevance of relative counts for contemporary biomedical data. While it is usually acknowledged that data produced by DNA sequencing instruments are relative \cite{ThomReview}, a number of arguments for the current dominance of absolute approaches have been put forward. We will discuss one of these arguments here: The constraint on the counts does not hold strictly, i.e., it is itself a fluctuating quantity \cite{Holmes}.\\
Counting the times $n_j$ a specific event $j$ occurs within a fixed time interval, under very general assumptions (i.e., independence of events from previous occurrences, fixed average rate of occurrence, no simultaneous occurrences), the resulting data will be distributed according to a Poisson distribution:
\begin{equation}
    p_P(n_j\mid\lambda_j)=\frac{\lambda_{j}^{n_{j}}}{n_{j}!}e^{-\lambda_{j}}.
\end{equation}
Here, $\lambda_j$ denotes the average occurrence rate\footnote{Recall that the $\lambda$ parameter coincides with the expected counts and also their variance. In practice, this could, e.g., be gene-transcriptional activities \cite{Nimwegen}.} of an event $j$. When considering $D$ such events now, and assuming they don't influence each other, we can write the overall probability of the $D$-dimensional vector of counts $\boldsymbol{n}$ simply as a product of $D$ such distributions.\\
Consider now a modification of this scenario where we observe these $D$ events taking place but instead of fixing a time interval, we will simply stop counting after we have observed $n$ events. The resulting distribution is a multinomial 
\begin{equation}
    p_n(\boldsymbol{n}\mid\boldsymbol{q})=\frac{n!}{\prod_{j=1}^Dn_j!}\prod_{j=1}^Dq_j^{n_j},\label{multinomial}
\end{equation}
where $\boldsymbol{q}=(q_j)_{j=1}^D$ is the vector of individual event probabilities\footnote{Note that we chose to put the auxiliary parameter $n$ as a subscript for a more compact notation}. The multinomial encodes a constraint on $\boldsymbol{n}$ that leads to a mutual dependence between the parts. In this sense, it models a composition of counts.\\
To see the connection between these two scenarios, let us come back to the independent Poisson distribution. It can be written as
\begin{multline}
    p_\mathrm{P}(\boldsymbol{n}\mid\boldsymbol{\lambda})=\prod_{j=1}^D\frac{\lambda_j^{n_j}}{n_j!}e^{-\lambda_j}
    \\=\frac{\lambda^n}{n!}e^{-\lambda}\frac{n!}{\prod_{j=1}^Dn_j!}\prod_{j=1}^D\left(\frac{\lambda_j}{\lambda}\right)^{n_j}
    =p_\mathrm{P}(n\mid\lambda)~p_n(\boldsymbol{n}\mid\boldsymbol{q}).\label{fac*multinomial}
\end{multline}
Here $\lambda$ denotes the sum over the components of $\boldsymbol{\lambda}$, and $\boldsymbol{q}=\boldsymbol{\lambda}/\lambda$. We see that the independent Poisson distributions factorize into a univariate Poisson of $n$ with parameter $\lambda$ as well as a multinomial distribution $p_n$ that has $n$ and $\boldsymbol{q}$ as parameters. This well-known relationship between the Poisson and the multinomial is interesting when discussing the argument against compositionality above. First we note that a variation in the constraining variable $n$ can only be used for a correct estimate of the rate parameters $\lambda_j$ of the $D$ Poisson processes if the overall rate $\lambda$ is exactly their sum. Modelling by a multinomial can thus be perfectly justified for a stochastic $n$ whose rate $\gamma$ is of no interest to the analyst because it is decoupled from the $\boldsymbol{\lambda}$, in the sense that $\gamma\ne\lambda$. For sequencing data, the constraint on $n$ is imposed by the capacity of the sequencing instrument while the variation in $n$ can be caused by other aspects of the protocol (e.g., the subsequent read mapping). The practical effects of the constraint are well documented \cite{TMM, Loven} and aren't invalidated by the stochastic nature of $n$.\\
{For an applicaton of the multinomial to single-cell sequencing data, see \cite{Irizarry}.} A pragmatic approach is taken in \cite{Nimwegen}, where it is acknowledged that the $q_j$, not the $\lambda_j$ should be the modelling objective, but (for practical reasons) their modelling is done by an independent Poisson that is reparametrized as $p_\mathrm{P}(\boldsymbol{n}\mid\lambda,\boldsymbol{q}$). The Poisson can serve as an approximation whenever there are no dominant parts for which $q_j$ becomes too large. The modelling gets complicated again as soon as co-variation of parts across samples are taken into account.
     
\subsection{Variation across samples, Bayes}

According to the Bayesian paradigm, probabilities are subjective in the sense that they quantify degrees of knowledge \cite{deFinetti}. This quantification involves both data and model parameters, and both can be arguments to probability functions. While we assume a fixed parameter when considering a single sample $\boldsymbol{n}$, it makes sense to let the parameter vary according to some distribution when considering many samples that were obtained under different conditions. This is typically the case when we have a data matrix where counts for $D$ variables (or compositional parts) indexed by the columns are collected in $N$ samples indexed by the rows.\\
As an example, consider the special case of the multinomial $p_n$. Our choice of the prior $\pi$ quantifying the probability of the parameter $\boldsymbol{q}$ will determine the functional form of the joint distribution and thus affect our ability to capture the variability across samples:
\begin{equation}
    p_n(\boldsymbol{n},\boldsymbol{q})=p_n(\boldsymbol{n}\mid\boldsymbol{q})\pi(\boldsymbol{q}).\label{joint}
\end{equation}
Integrating the joint probability\footnote{While we use the convention to denote it by the same symbol as the likelihood, this is generally not a multinomial.} over the parameter $\boldsymbol{q}$ would leave us again with $\boldsymbol{n}$ as the only argument. The resulting marginal distribution will depend on the hyperparameters of the prior (which we left out in the formula above)\footnote{An example that concerns much of the current modelling of sequencing data is going from the Poisson distribution to the (overdispersed) negative binomial distribution when integrating out the original $\lambda_j$ parameter with a conjugate gamma prior.}. If we divide (\ref{joint}) by it, we renormalize and obtain the posterior probability of the parameter $\boldsymbol{q}$, giving us Bayes' theorem.\\
An excellent choice for $\pi$ would be a $D-1$-dimensional multivariate normal of the log-ratios $\log(q_i/q_D)$. This allows for a compositional modelling of the second-order interactions between parts that captures the over-dispersion often observed in real-world data \cite{Billheimer, Li}. While this logistic-normal multinomial model has no analytic solution, Markov-Chain Monte Carlo can be used, like in a recent application to differential association networks in microbiome data \cite{mdine}. Note that the interest is now in the hyperparameters of the prior, especially in the covariance matrix of the log-ratios of $\boldsymbol{q}$.\\
A less realistic but more tractable solution is obtained when simply choosing the conjugate prior to the multinomial, i.e., the Dirichlet distribution. While we will later describe it in more detail, let us here point out that this model implies that all interaction between parts comes from the constraint that counts have to add to $n$. It is thus the model with the greatest degree of independence that can be achieved for compositions \cite{AitchisonBook}. 

\subsection{Dual Coordinates for Count Distributions}

We have recently proposed to treat compositional data with the methods of information geometry \cite{ErbAy}. 
{The fact that the geometric structure of the discrete probability simplex can be exploited for the analysis of compositional data has been observed before, e.g. \cite{Sun}. Compositions $\boldsymbol{q}$ can be described as categorical distributions that live on a finite dimensional open}\footnote{{This is a technical requirement so we can use logarithms. More often than not, compositional data will fall on a closed simplex \cite{Greenacre2021}.}} simplex 
\begin{equation}
    \mathcal{S}^D=\left\{(q_1,\dots,q_D)^T\in\mathbb{R}^D:q_i>0,i=1,\dots,D,\sum_i^Dq_i=1\right\}.
\end{equation}
{The finite version of information geometry contains already all its important concepts but often provides a more intuitive approach, see \cite{Amari,GzylNielsen}. For a comprehensive treatment of the finite case, see Chapter 2 of \cite{Nihat}.  
We are now showing a concrete example of an application to CoDA that slightly extends our framework in \cite{ErbAy} to deal with relative count data.}\\
To briefly recapitulate, we start from the two natural coordinate systems used in information geometry: the expectation parameters $\boldsymbol{\eta}$ (whose components carry lower indices) and the exponential parameters $\boldsymbol{\theta}$ (with upper indices). Consider again the case where the occurrence of $D$ discrete events is encoded by a random variable $R=r\in\{1,\dots,D\}$ with occurrence probabilities $\boldsymbol{q}$. The $D-1$-dimensional vector of expectation parameters $\boldsymbol{\eta}$ consists simply of those probabilities that can vary freely (while all of them  have to sum to 1). The probability of an event in terms of $\boldsymbol{\eta}$ can then be written as
\begin{equation} \label{etapara}
    p(r\mid\boldsymbol{\eta}) =
    \left\{
      \begin{array}{c@{\quad}l}
         \eta_r & \mbox{if $r \leq D-1$,} \\
         1 - \sum_{i = 1}^{D-1} \eta_i 
         & \mbox{if $r = D$,}
      \end{array}
    \right.  \qquad r = 1,\dots,D.
\end{equation}
Alternatively, this distribution can be parametrized using what is known as the alr-transformation in CoDA:
\begin{equation}
    \theta^j=\log\frac{q_j}{q_D},~~~j=1,\dots,D-1.\label{theta}
\end{equation}
Note that we are not (as often done in CoDA) log-ratio transforming the data themselves, but their underlying parameters $\boldsymbol{q}$. With this, we can write our distribution in the form  
\begin{equation} 
    p(r\mid\boldsymbol{\theta}) =  \mathrm{exp}\left(\sum_{k=1}^{D-1}\theta^k {\mathbbm 1}_k(r) -\psi(\boldsymbol{\theta})\right), \qquad 
    r = 1,\dots,D,
    \label{thetadist2}
\end{equation}
where ${\mathbbm 1}_k(r) = 1$ if $r = k$, and ${\mathbbm 1}_k(r) = 0$ otherwise.   
The function $\psi$ ensures normalization and is known as the free energy. It is given by
\begin{equation}
    \psi(\boldsymbol{\theta})=\log\left(1+\sum_{i=1}^{D-1}e^{\theta^i}\right)=-\log q_D.\label{psi}
\end{equation}
How do we get from a single outcome $r$ to the multinomial counts $\boldsymbol{n}$? Let us first consider $n$ outcomes $\boldsymbol{r}=(r_1,\dots,r_n)$. Their probability is simply the product over (\ref{thetadist2}):
\begin{eqnarray} 
    p(\boldsymbol{r}\mid n,\boldsymbol{\theta}) &=&\prod_{i=1}^np(r_i\mid\boldsymbol{\theta})\nonumber\\
    &=&\exp\sum_{i=1}^{n}\left(\sum_{k=1}^{D-1}\theta^k{\mathbbm 1}_k(r_i) -\psi(\boldsymbol{\theta})\right),\nonumber\\
    &=&\exp\left(\sum_{k=1}^{D-1}\theta^kn_k(\boldsymbol{r})-n\psi(\boldsymbol{\theta})\right),
    \label{expo2}
\end{eqnarray}
where $n_k(\boldsymbol{r}):=\sum_{i=1}^{n}{\mathbbm 1}_k(r_i)$. This latter expression encodes the 
$D$ components of our relative counts $\boldsymbol{n}$. To obtain their probability of occurrence, we note that many outcomes $\boldsymbol{r}$ lead to the same outcomes of counts. Counting these leads to a factor given by the multinomial coefficient:
\begin{equation}
    p_0(\boldsymbol{n}\mid n)=\frac{n!}{n_1!\dots n_D!}={n\choose n_1\dots n_D}.
\end{equation}
With this base measure, we can finally write our multinomial (\ref{multinomial}) in form of an exponential family
\begin{equation}
    p_n(\boldsymbol{n}\mid\boldsymbol{\theta})=p_0(\boldsymbol{n}\mid n)~\mathrm{exp}\left(\sum_{k=1}^{D-1}\theta^k n_k -n\psi(\boldsymbol{\theta})\right).\label{expo_mult}
\end{equation}
We see that the exponential coordinates remain the same regardless of the number of observations. It is often convenient to drop the base measure and, changing the random variable, resort to the expression (\ref{expo2}). Also, as we can see from (\ref{expo_mult}), to obtain the multi-event versions of $\boldsymbol{\eta}$ and $\psi(\boldsymbol{\theta})$, we just need to multiply by $n$. Due to the Legendre duality of the natural coordinates, we can obtain the multi-event expectation coordinates by taking partial derivatives 
\begin{equation}
    n\eta_j=\frac{\partial}{\partial\theta^j}n\psi(\boldsymbol{\theta})=\mathbbm{E}_{p_n}(n_j)=nq_j,\qquad j=1,\dots,D-1.\label{eta}
\end{equation}
 Finally, the potential that is dual to the multi-event free energy $n\psi(\boldsymbol{\theta})$, i.e., the negative Shannon entropy of (\ref{expo2}), is given by $n\phi(\boldsymbol{\eta})$, where
\begin{equation}
    \phi(\boldsymbol{\eta})=\sum_{k=1}^{D-1}\eta_k\log\eta_k+\left(1-\sum_{k=1}^{D-1}\eta_k\right)\log\left(1-\sum_{k=1}^{D-1}\eta_k\right).\label{phi}
\end{equation}

\subsection{Parameter Divergence from Observed Points}

In the previous section, we have derived expressions for probabilities of data given some model parameters. These parameters happen to be compositions, and as such they can be depicted as points in a simplex. When normalizing a sample of count data by their total, we can also represent it as a so-called {\it observed point} \cite{Amari} in the simplex: 
\begin{equation}
    \hat{\boldsymbol{q}}=\left(\frac{n_1}{n},\dots,\frac{n_D}{n}\right)^T.\label{q_emp}
\end{equation}
This is the empirical estimate of the parameter $\boldsymbol{q}$. {The empirical estimate} is also known as the {\it type} of a sequence $\boldsymbol{r}$ of independent random variables. Our dual coordinates associated with the observed point are
\begin{eqnarray}
    \hat{\boldsymbol{\theta}}&=&\left(\log\frac{n_1}{n_D},\dots,\log\frac{n_{D-1}}{n_D}\right)^T,\label{theta_emp}\\
    n\hat{\boldsymbol{\eta}}&=&\left(n_1,\dots,n_{D-1}\right)^T.\label{eta_emp}
\end{eqnarray}
One of the fundamental results of the method of types  (e.g., \cite{CoverThomas}) is an equality relating the true distribution to the observed point:
\begin{equation}
    p(\boldsymbol{r}\mid n,\boldsymbol{\theta}) =\exp\left(n\phi(\hat{\boldsymbol{\eta}})-n D_\phi(\hat{\boldsymbol{q}}\mid\mid\boldsymbol{q})\right),\label{MOT}
\end{equation}
where 
\begin{equation}
    D_\phi(\hat{\boldsymbol{q}}\mid\mid\boldsymbol{q})=\sum_{j=1}^{D}\frac{n_j}{n}\log\frac{n_j}{n q_j}
\end{equation}
is the relative entropy, or Kullback-Leibler (KL) divergence, between the empirical and the true parameter compositions. The expression (\ref{MOT}) can be easily derived by simple algebraic rearrangement of (\ref{expo2}) using the expressions for $\phi$ and $D_\phi$. With (\ref{MOT}), it is clear that we can write {the multi-event version of our divergence as
\begin{equation}
    nD_\phi(\hat{\boldsymbol{q}}\mid\mid\boldsymbol{q})=n\phi(\hat{\boldsymbol{\eta}})-\log p(\boldsymbol{r}\mid n,\boldsymbol{\theta}).\label{maxlikeli}
\end{equation}
}As the first term does not depend on $\boldsymbol{\theta}$, this shows why taking the maximum of the likelihood $p(\boldsymbol{r}\mid n,\boldsymbol{\theta}))$ over $\boldsymbol{\theta}$ is equivalent to minimizing the KL-divergence between the estimated and the true parameter composition.\\
More general relationships of this kind can be derived from a fundamental information-geometric equality {that is due to the Legendre duality between $\boldsymbol{\psi}$ and $\boldsymbol{\phi}$:}
\begin{equation}
    D_\phi(\hat{\boldsymbol{q}}\mid\mid\boldsymbol{q})=\phi(\hat{\boldsymbol{\eta}})+\psi(\boldsymbol{\theta})-\boldsymbol{\theta}^T\hat{\boldsymbol{\eta}}\label{KL}.
\end{equation}
Minimizing a dissimilarity between distributions can be understood as a projection. Here we project the observed point onto the manifold of distributions parametrized by $\boldsymbol{\theta}$. In information geometry, this minimization of the KL-divergence is known under the name of $m$-projection{, see \cite{Amari}. In section \ref{generalEst}, we will show a result that is more general than (\ref{maxlikeli}) in the sense that it does not only hold for the likelihood but also for prior and posterior probability.}

\subsection{Posterior Probability of the Parameter}\label{PostProb}

For the Bayesian estimation a parameter we have to construct a posterior distribution of the parameter that also takes into account its prior distribution $\pi$, which itself can depend on a vector of hyperparameters $\boldsymbol{\alpha}$. {For a review of Bayesian inference for categorical data see \cite{Agresti}.}
The posterior probability density of the parameter in terms of the exponential parameter $\boldsymbol{\theta}$ is
\begin{equation}
    p(\boldsymbol{\theta}\mid\boldsymbol{r},n,\boldsymbol{\alpha})=\frac{p(\boldsymbol{r}\mid n,\boldsymbol{\theta})\pi(\boldsymbol{\theta}\mid\boldsymbol{\alpha})}{\int d\boldsymbol{\theta}^\prime p(\boldsymbol{r}\mid n,\boldsymbol{\theta}^\prime)\pi(\boldsymbol{\theta}^\prime\mid\boldsymbol{\alpha})}.\label{Bayes}
\end{equation}
Instead of maximizing the likelihood over $\boldsymbol{\theta}$, we can now maximize the posterior to obtain the best parameter estimate\footnote{Alternatively, we could take the expectation value of $\boldsymbol{\theta}$ with respect to its posterior.}. 
Inserting (\ref{expo2}), the posterior (\ref{Bayes}) evaluates to
\begin{equation}
    p(\boldsymbol{\theta}\mid\boldsymbol{r},n,\boldsymbol{\alpha})=\pi(\boldsymbol{\theta}\mid\boldsymbol{\alpha})\exp\left(\sum_{k=1}^{D-1}\theta^k n_k(\boldsymbol{r}) -n\psi(\boldsymbol{\theta})-\log p(\boldsymbol{r}\mid\boldsymbol{\alpha})\right).
    \label{posterior}
\end{equation}
where $p(\boldsymbol{r}\mid\boldsymbol{\alpha})$ is the normalizing integral in the denominator of (\ref{Bayes}). Seeing this as an exponential family, we note that the parameter and the random variables have exchanged their roles. The prior can be written as a new base measure now, while the new free energy is given by $\log p(\boldsymbol{r}\mid\boldsymbol{\alpha})$.\footnote{{To explain the extra term $-n\psi(\boldsymbol{\theta})$ in this picture, $n$ and $-\psi(\boldsymbol{\theta})$ can be considered extra components of the vectors $\boldsymbol{n}$ and $\boldsymbol{\theta}$, respectively.}}\\
A prior that has the same functional form as the resulting posterior is called a conjugate prior. Using a conjugate prior makes closed-form solutions of the posterior possible. The general form of the conjugate prior for an exponential family is well known \cite{Diaconis}, but it is instructive to obtain it as follows. We copy the functional form of (\ref{posterior}) and obtain a $D$-parameter conjugate prior as
\begin{equation}
    \pi(\boldsymbol{\theta}\mid\boldsymbol{\alpha})=\pi_0(\boldsymbol{\theta})\exp\left(\sum_{k=1}^{D-1}\theta^kf_k(\boldsymbol{\alpha})-\left[\sum_{k=1}^Df_k(\boldsymbol{\alpha})\right]\psi(\boldsymbol{\theta})-\chi(\boldsymbol{\alpha})\right),\label{conprior}
\end{equation}
where $\pi_0$ is a base measure, $f_k$ is a sufficient statistic of the $k$-th hyperparameter, and $\chi$ the normalization. With this, the posterior (\ref{posterior}) becomes
\begin{multline}
    p(\boldsymbol{\theta}\mid\boldsymbol{r},n,\boldsymbol{\alpha})=\pi_0(\boldsymbol{\theta})\times\\
    \exp\left(\sum_{k=1}^{D-1}\theta^k\left(n_k(\boldsymbol{r})+f_k(\boldsymbol{\alpha})\right)-\left[n+\sum_{k=1}^Df_k(\boldsymbol{\alpha})\right]\psi(\boldsymbol{\theta})-\chi(\boldsymbol{\alpha})-\log{p(\boldsymbol{r}\mid\boldsymbol{\alpha})}\right).\label{posterior2}
\end{multline}
In our {categorical case it is well known \cite{Agresti}} that the conjugate prior is a Dirichlet distribution with parameters $\boldsymbol{\alpha}$. The expressions involved evaluate to 
\begin{eqnarray}
    f_k(\boldsymbol{\alpha})&=&\alpha_k,\label{suffstat}\\
      \pi_0(\boldsymbol{\theta})&=&1,\\
      \chi(\boldsymbol{\alpha})&=&\log B(\boldsymbol{\alpha}),\label{betafunc}\\
      p(\boldsymbol{r}\mid\boldsymbol{\alpha})&=&\frac{B\left((n_k(\boldsymbol{r})+\alpha_k)_{k=1}^D\right)}{B(\boldsymbol{\alpha})},\label{dataprob}
\end{eqnarray}
where $B$ denotes the multivariate beta function. (For clarity, we give a short derivation for $p(\boldsymbol{r}\mid\boldsymbol{\alpha})$ in the Appendix.) With these expressions, the posterior simplifies to
\begin{multline}
    p(\boldsymbol{\theta}\mid\boldsymbol{r},n,\boldsymbol{\alpha})=\\
    \exp\left(\sum_{k=1}^{D-1}\theta^k\left(n_k(\boldsymbol{r})+\alpha_k)\right)-\left[n+\sum_{k=1}^D\alpha_k\right]\psi(\boldsymbol{\theta})-\log B\left(\boldsymbol{n}(\boldsymbol{r})+\boldsymbol{\alpha}\right)\right).\label{posterior3}
\end{multline}
We can see here the {widely-used result} that the posterior is obtained from the likelihood by simply adding the conjugate prior parameters as pseudo counts to the respective event counts and then renormalizing.

\subsection{Parameter Divergence from General Estimators}\label{generalEst}

{The similarity between the likelihood and our expression for the posterior} suggests that we can maximize the posterior similarly to the likelihood by minimizing a certain KL-divergence. Indeed, {the following proposition shows that maximizing prior, likelihood, or posterior always corresponds to a minimization of KL-divergence between a suitable estimator and $\boldsymbol{q}$:}\\ 

\begin{prop}\label{KLMAP}
{
Let $\boldsymbol{q}$ be a parameter of probabilities with exponential coordinates $\boldsymbol{\theta}$ via $p(r\mid\boldsymbol{\theta})$ with free energy $\psi(\boldsymbol{\theta})$ as defined in (\ref{theta})-(\ref{psi}). Further, let the function $f:\mathcal{S}^D\times\mathbb{R}_+\times\mathbb{R}^{D-1}\to\mathbb{R}_+$ be given by 
\begin{equation}
f(\tilde{\boldsymbol{q}},\tilde{n},\boldsymbol{\theta})=Z(\tilde{n},\tilde{\boldsymbol{q}})~\mathrm{exp}\left\{\tilde{n}\left(\boldsymbol{\theta}^T\tilde{\boldsymbol{\eta}}-\psi(\boldsymbol{\theta})\right)\right\},\nonumber
\end{equation}
where $\tilde{\boldsymbol{q}}$ is an estimator of $\boldsymbol{q}$ with expectation coordinates $\tilde{\boldsymbol{\eta}}$, $\tilde{n}$ denotes a positive real, and $Z$ a positive function. We then have
\begin{equation}
\tilde{n}D_\phi(\tilde{\boldsymbol{q}}\mid\mid \boldsymbol{q})=\tilde{n}\phi(\tilde{\boldsymbol{\eta}})+Z(\tilde{n},\tilde{\boldsymbol{q}})-\log f(\tilde{\boldsymbol{q}},\tilde{n},\boldsymbol{\theta}),\nonumber
\end{equation}
with $\phi$ the Lagrange dual to $\psi$ as defined in (\ref{phi}) and $D_\phi$ the KL-divergence.}
\end{prop}
{The proof makes use of (\ref{KL}) and otherwise consists in a simple rearrangement of terms (see Appendix).}
\begin{cor}
{
Maximization of $\log f(\tilde{\boldsymbol{q}},\tilde{n},\boldsymbol{\theta})$ as a function of $\boldsymbol{\theta}$ minimizes $D_\phi(\tilde{\boldsymbol{q}}\mid\mid \boldsymbol{q})$ as a function of $\boldsymbol{q}$.
}
\end{cor}

{This is clear because the other (data-dependent) terms do not depend on the parameter.}\\
\begin{ex}
{Shrinkage estimator:}
\end{ex}
 We use as our estimator {$\tilde{\boldsymbol{q}}$} the expected value of $\boldsymbol{q}$ under the posterior (\ref{posterior3}), the so-called \emph{shrinkage estimator} $\hat{\boldsymbol{q}}^\mathrm{sh}$
\begin{equation}
    \tilde{\boldsymbol{q}}=\hat{\boldsymbol{q}}^\mathrm{sh}:=\mathbb{E}_{\boldsymbol{\theta}}(\boldsymbol{q}\mid \boldsymbol{r},n,\boldsymbol{\alpha})=\frac{\boldsymbol{n}+\boldsymbol{\alpha}}{n+\sum_{k=1}^D\alpha_k},\label{shrink1}
\end{equation}
{and set $\tilde{n}=\hat{n}:=n+\sum_{k=1}^D\alpha_k$.  This allows us to reparametrize the posterior in the required form 
\begin{equation}
    p(\boldsymbol{\theta}\mid\hat{\boldsymbol{q}}^\mathrm{sh},\hat{n})=\exp\left(\hat{n}\left[\sum_{k=1}^{D-1}\theta^k\hat{q}_k^\mathrm{sh}-\psi(\boldsymbol{\theta})\right]-\log B\left(\hat{n}\hat{\boldsymbol{q}}^\mathrm{sh}\right)\right),\label{posteriorSH}
\end{equation}
and thus $f(\tilde{\boldsymbol{q}},\tilde{n},\boldsymbol{\theta})=p(\boldsymbol{\theta}\mid\boldsymbol{r},n,\boldsymbol{\alpha})$ and $Z(\tilde{n},\tilde{\boldsymbol{q}})=1/B(\hat{n}\hat{\boldsymbol{q}}^\mathrm{sh})$. With this, the proposition gives} 
\begin{equation}
    \hat{n}D_\phi(\hat{\boldsymbol{q}}^\mathrm{sh}\mid\mid\boldsymbol{q})=\hat{n}\phi(\hat{\boldsymbol{\eta}}^\mathrm{sh})-\log B(\hat{n}\hat{\boldsymbol{q}}^\mathrm{sh})-\log p(\boldsymbol{\theta}\mid\hat{\boldsymbol{q}}^\mathrm{sh},\hat{n}).\label{maxpost}
\end{equation}
Thus finding the $\boldsymbol{\theta}$ that maximizes the posterior is equivalent to minimizing the KL-divergence between the shrinkage estimator and the true parameter $\boldsymbol{q}$. 

\begin{ex}
{Empirical estimator:}
\end{ex}
{The empirical estimator of the multinomial distribution is a straightforward application: $\tilde{\boldsymbol{q}}=\hat{\boldsymbol{q}}:=\boldsymbol{n}/n$, $\tilde{n}=n$, and $f(\tilde{\boldsymbol{q}},\tilde{n},\boldsymbol{\theta})= p_n(\boldsymbol{n}\mid\boldsymbol{\theta})$ as given by (\ref{expo_mult}), so $Z(\tilde{n},\tilde{\boldsymbol{q}})$ is the multinomial coeffcient $p_0(\boldsymbol{n}\mid n)$. The proposition gives (\ref{maxlikeli}) with an additional subtraction of the $\log p_0$ term.}\\

{Clearly, another example consists in maximizing the prior probability of $\boldsymbol{\theta}$ to minimize the divergence between $\boldsymbol{\alpha}/\sum_k\alpha_k$ and $\boldsymbol{q}$. In section \ref{application} we will define another version of the shrinkage estimator, which will provide us with yet another application of the proposition. Note that $f(\tilde{\boldsymbol{q}},\tilde{n},\boldsymbol{\theta})$ has the general form of a conjugate prior of an exponential family, so Proposition 1 holds for exponential families in general. A more general treatment than the one presented here can be found in \cite{AgarwalDaume}.}

\subsection{Decision-Theoretic Risk}\label{decision}

Decision theory (e.g., \cite{Berger}) provides a foundational framework for statistics. While it is closely linked with Bayesian analysis, it can also be formulated from a frequentist point of view. In any case, it implies the construction of a loss function that incorporates statistical knowledge in order to quantify the risk of a wrong decision. Such a loss function $L$ has the ``true state of nature" and an action (based on some knowledge) as its arguments. Perhaps the most important example for these arguments would be the true parameter $\boldsymbol{q}$ of a distribution and some estimator $\hat{\boldsymbol{q}}$, where the latter would be identified with the action based on it. Given some loss $L(\boldsymbol{q},\hat{\boldsymbol{q}})$, the risk we incur when basing our decision on the estimator is then some expected value
\begin{equation}
    R(\hat{\boldsymbol{q}})=\mathbbm{E}L(\boldsymbol{q},\hat{\boldsymbol{q}}).
\end{equation}
Bayesian and frequentist schools disagree on the type of expectation that should be taken here. While for the Bayesian the expectation is taken with respect to the posterior probability\footnote{In a data-free context, it can also be taken with respect to the prior probability.} of the parameter $\boldsymbol{q}$, the frequentist averages over all instances of the random variables (which follow a distribution parametrized by $\boldsymbol{q}$)\footnote{{An example of such a risk function is $D_\phi(\hat{\boldsymbol{q}}\mid\mid\boldsymbol{q})$.}}. As a consequence, the risk remains a function of $\boldsymbol{q}$. A frequentist then calls an estimator $\hat{\boldsymbol{q}}_1$ $R$-better than $\hat{\boldsymbol{q}}_2$ when  $R_{\boldsymbol{q}}(\hat{\boldsymbol{q}}_1)\le R_{\boldsymbol{q}}(\hat{\boldsymbol{q}}_2)$ for all $\boldsymbol{q}$, with strict inequality for some of them. An estimator is called {\it inadmissible} if there exists an $R$-better estimator.\\
Often, for pragmatic reasons, a quadratic loss leading to a mean squared error (MSE) risk function is assumed. Beside its simplicity, one benefit is that for unbiased estimators, the (frequentist) risk is simply the variance of the estimator:
\begin{equation}
    R_{\boldsymbol{q}}(\hat{\boldsymbol{q}})=\mathbbm{E}\left[(\hat{\boldsymbol{q}}-\boldsymbol{q})^2\right]=\sum_{j=1}^{D}\left[\mathrm{var}(\hat{q}_j-q_j)+\mathbbm{E}^2(\hat{q}_j-q_j)\right]=\sum_{j=1}^{D}\mathrm{var}(\hat{q}_j).\label{MSE1}
\end{equation}
Here, the bias-variance decomposition of the MSE was used, and the last equality follows from the facts that $q_j$ is not stochastic and that the bias $\mathbbm{E}\left[\hat{\boldsymbol{q}}-\boldsymbol{q}\right]$ vanishes. Note that here we do not have to know the true value of $\boldsymbol{q}$ to evaluate its risk because in practice, to evaluate the variance of the estimator, its empirical estimate is used. As an example, for the empirical estimator (\ref{q_emp}), the variance components would be estimated by $\hat{q}_j(1-\hat{q}_j)/(n-1)$.

\subsection{James-Stein shrinkage and regularization}\label{shrinkage}

The empirical estimator $\hat{\boldsymbol{q}}$ {is (unlike the empirical estimator of the multivariate normal mean) known to be admissible under quadratic loss \cite{Johnson}, so there is no "Stein effect"  \cite{Stein} for the multinomial. While the Bayesian estimator (\ref{shrink1}) isn't uniformly better than the empirical estimator for all parameter values,\footnote{An example where the empirical estimator gives a better value for $q_j$ is the case where $n_j=0$ and the prior value of the Bayesian estimator is further away from $q_j$ than $q_j$ is from zero.} its flattening of the data can result in much smaller mean squared error than with the empirical estimate. This will be made plausible in the following.} Let us rewrite (\ref{shrink1}) as a convex combination
\begin{equation}
    \hat{\boldsymbol{q}}^\mathrm{sh}=\lambda\boldsymbol{\tau}+(1-\lambda)\hat{\boldsymbol{q}}\label{shrink2}
\end{equation}
between the {\it target distribution} $\boldsymbol{\tau}$ and the empirical estimator $\hat{\boldsymbol{q}}$. That this is equivalent to (\ref{shrink1}) can be seen when defining
\begin{eqnarray}
    \lambda&:=&\frac{\sum_{k=1}^D\alpha_k}{n+\sum_{k=1}^D\alpha_k},\label{lambda}\\
    \tau_j&:=&\frac{\alpha_j}{\sum_{k=1}^D\alpha_k},\qquad j=1,\dots,D.\label{tau}
\end{eqnarray}
$\hat{\boldsymbol{q}}^\mathrm{sh}$ is called a James-Stein type \cite{JamesStein} shrinkage estimator of $\boldsymbol{q}$, see also \cite{EfronMorris} as well as the discussion in \cite{HausserStrimmer}. Choosing the maximum-entropy target, i.e., the equidistribution $\tau_j=1/D$ for all $j=1,\dots,D$, the target term can be understood as a regularization of the empirical estimator.\\
{Remember that $\hat{\boldsymbol{q}}^\mathrm{sh}$ is the posterior expected value of $\boldsymbol{q}$. The fact that the posterior expected value of a random variable is a linear function of its empirical estimate is equivalent to the use of a conjugate prior. This is a result that holds for exponential families in general \cite{Diaconis}.\\
This linearity is helpful for evaluating the accuracy of the shrinkage estimator, again using the expected quadratic loss as a risk function. We shall give a result that is slightly more general than necessary for this estimator because we will again need it in section \ref{application}:} 
\begin{prop}\label{LedoitWolf}
{Let $f_j$, $j=1,\dots,D$ be the components of a function $f:\mathcal{S}^D\to\mathbb{R}^D$ acting on a vector of probabilities. Let $\boldsymbol{\tau}$ be a $D$-dimensional probability parameter and $\hat{\boldsymbol{q}}$ the multinomial empirical estimator. Then, for $0\le\lambda\le1$, the convexly combined estimator $f(\tilde{\boldsymbol{q}})$ of $f(\boldsymbol{q})$ given by its components
\begin{equation}
    f_j(\tilde{\boldsymbol{q}}):=\lambda f_j(\boldsymbol{\tau})+(1-\lambda)f_j(\hat{\boldsymbol{q}}),\qquad j=1,\dots,D\nonumber
\end{equation}
(i) has a quadratic risk with respect to $f(\boldsymbol{q})$ given by
\begin{equation}
    R_{\boldsymbol{q}}(\tilde{\boldsymbol{q}})=(1-\lambda)^2\sum_{j=1}^D\mathrm{var}\big(f_j(\hat{\boldsymbol{q}})\big)+\sum_{j=1}^D\bigg[\mathbb{E}f_j(\hat{\boldsymbol{q}})-f_j(\boldsymbol{q})-\lambda\big(\mathbb{E}f_j(\hat{\boldsymbol{q}})-f_j(\boldsymbol{\tau})\big)\bigg]^2.\nonumber
\end{equation}
(ii) The minimum risk is attained for
\begin{equation}
    \lambda^*=\frac{\sum_{j=1}^D\bigg[\mathrm{var}\big(f_j(\hat{\boldsymbol{q}})\big)+\big(\mathbb{E}f_j(\hat{\boldsymbol{q}})-f_j(\boldsymbol{q})\big)\big(\mathbb{E}f_j(\hat{\boldsymbol{q}})-f_j(\boldsymbol{\tau})\big)\bigg]}{\sum_{j=1}^D\mathbb{E}\big[f_j(\hat{\boldsymbol{q}})-f_j(\boldsymbol{\tau})\big]^2}.\nonumber
\end{equation}
}
\end{prop}
{The proof is provided in the Appendix. This is a slight modification of the lemma shown in \cite{LedoitWolf}, see also the derivation in \cite{SchaeferStrimmer} and the application to the multinomial in \cite{HausserStrimmer}. To apply the proposition to $\hat{\boldsymbol{q}}^\mathrm{sh}$, we observe that $f_j$ simply corresponds to taking the $j$-th component and simplifications occur because the bias of $\hat{\boldsymbol{q}}$ vanishes: $\mathbb{E}f_j(\hat{\boldsymbol{q}})-f_j(\boldsymbol{q})=\mathbb{E}\hat{q}_j-q_j=0$. We obtain}  
\begin{equation}
    R_{\boldsymbol{q}}(\hat{\boldsymbol{q}}^\mathrm{sh})=(1-\lambda)^2\sum_{j=1}^{D}\mathrm{var}(\hat{q}_j)+\lambda^2\sum_{j=1}^{D}\mathbbm{E}^2(\hat{q}_j-\tau_j),\label{cost}
\end{equation}
with minimum risk at
\begin{equation}
    \lambda^*=\frac{\sum_{j=1}^{D}\mathrm{var}(\hat{q}_j)}{\sum_{j=1}^{D}\mathbbm{E}\left[(\hat{q}_j-\tau_j)^2\right]}.\label{min}
\end{equation}
We can see that the risk function is a weighted average over the risk of the empirical estimator and an additional term that punishes expected difference from the target. {Tuning the size of $\lambda$, we can trade off the bias of the target against the variance of the empirical estimate to obtain a smaller risk than (\ref{MSE1}). Estimators based on small sample data will generalize better to new data when flattening the data to a well-specified extent using an uninformative, maximum-entropy model. The amount of flattening depends on the data at hand and is} optimized via the weight $\lambda$ of the target. Note that the relationships (\ref{lambda}) and (\ref{tau}) imply that this is similar to an empirical Bayes procedure where we tune the size of the pseudocounts $\alpha_j$ and by this, adjust the a-priori sample size $\sum\alpha_k=n\lambda/(1-\lambda)$. To evaluate (\ref{min}), the empirical estimates for variance and expectation are used in practice.

\subsection{Power-Transformed Compositions and their Euclidean Distance in Ordination}
Power transformations \cite{Power} have traditionally been applied to data in order to fulfill certain distributional assumptions. For instance, a suitable power transformation can reduce skew so data appear approximately normal. In the case of Poisson counts, where variance equals the mean, the square root transformation is a common choice to ``stabilize" the variance (i.e., make it approximately constant independently of the mean). More generally, power transformations can appear through the link functions of generalized linear models \cite{Biplots} and then enable a fit of the data to a true underlying distribution.\\
Methods for dimension reduction and data visualization (a.k.a.\ \emph{ordination}) such as Principal Component Analysis (PCA) often use some version of Euclidean distance between multivariate samples: 
\begin{equation}
    d^2(\hat{\boldsymbol{q}}_1,\hat{\boldsymbol{q}}_2)=\sum_{j=1}^D\omega_j\left(\hat{q}_{1j}-\hat{q}_{2j}\right)^2,\label{chisquare}
\end{equation}
where the $\omega_i$ are suitable weights. Here, for the data, we used the empirical parameter estimates of the count distribution $\hat{\boldsymbol{q}}$ instead of the counts $\boldsymbol{n}$ themselves. In the case of relative counts, where the total of each sample is not of direct interest, this seems a good idea because we want to visualize the ``shape" of the data without their ``size" \cite{Greenacre2017}. There are two main ordination methods that are relational in the sense that they visualize shape only \cite{Biplots}, Correspondence Analysis (CA) and log-ratio analysis (LRA). CA uses a weighting scheme that involves row and column totals of the data matrix. In this way, it takes into account the data size indirectly to account for the precision of the shape estimates. LRA, in contrast, is a PCA of data that are log-transformed and double-centred. Here, relationships between parts remain invariant under taking subsets of the data,\footnote{This property is known as subcompositional coherence.} and it is better suited for true compositions. It was shown \cite{Greenacre2010} that via the following limit of the Box-Cox family \cite{BoxCox} of power transformations
\begin{equation}
    \lim_{\beta\to0}\frac{x^\beta-1}{\beta}=\log(x),\label{BoxCox}
\end{equation}
CA on power-transformed data converges to LRA. CA and LRA are thus special cases of a more general family of ordination methods. To make this more precise in the case of unweighted LRA, consider the following transformation of our empirical estimates: 
\begin{equation}
    f_\beta(\hat{\boldsymbol{q}})=\left(\frac{\hat{q}_1^\beta}{\sum_{k=1}^D\hat{q}_k^\beta},\dots,\frac{\hat{q}_D^\beta}{\sum_{k=1}^D\hat{q}_k^\beta}\right)^T.\label{powering}
\end{equation}
When now using uniform weights $\omega_j=D^2$, the limit
\begin{equation}
    \lim_{\beta\to0}\frac{1}{\beta^2}d^2\left(f_\beta(\hat{\boldsymbol{q}}_1),f_\beta(\hat{\boldsymbol{q}}_2)\right)
\end{equation}
is the squared Aitchison distance
\begin{equation}
    d^2_A(\hat{\boldsymbol{q}}_1,\hat{\boldsymbol{q}}_2)=\frac{1}{D}\sum_{i=1}^D\sum_{j<i}\left(\log\frac{\hat{q}_{1i}}{\hat{q}_{1j}}-\log\frac{\hat{q}_{2i}}{\hat{q}_{2j}}\right)^2\label{Aitchison}
\end{equation}
(see \cite{ErbAy} for a proof). Aitchison (or log-ratio) distance is the metric underlying LRA.  Using the transformation $f_\beta$ before evaluating Euclidean distance induces a parametrized class of distance measures that include the ones used in CA ($\beta=1$) and LRA ($\beta=0$) as special cases\footnote{Note that the row weights are assumed to be uniform for the special case of compositional data.}. When using finite, ``small  enough" values of the power parameter $\beta$, the subcompositional coherence of LRA remains approximately satisfied while there is no need for zero imputation (as CA does not involve logarithms).
One can obtain an optimal value of the power parameter in the sense that it maximizes the Procrustes correlation between the log-ratio transformed data (using zero imputation) and the  coordinates from the power-transformed CA (keeping the zeros) \cite{Reappraisal}.

\section{Exponential Shrinkage}\label{application}

{
In this section we want to define and test an estimator based on the power transformation (\ref{powering}). The justification of this estimator comes from a formal analogy with $\hat{\boldsymbol{q}}^\mathrm{sh}$. This analogy is more apparent when introducing the generalized notions of addition (a.k.a.\ \emph{perturbation}) and scalar multiplication (a.k.a.\ \emph{powering}) that equip the simplex with a linear structure. For  $\boldsymbol{q}, \boldsymbol{p}\in\mathcal{S}^D$, and some $\beta\in\mathbb{R}$, they are defined as the vectors
\begin{eqnarray}
\boldsymbol{q}\oplus\boldsymbol{p}&:=&\mathcal{C}(q_1p_1,\dots,q_Dp_D)^T,  \label{pert} \\
\beta\odot\boldsymbol{q}&:=&\mathcal{C}(q_1^\beta,\dots,q_D^\beta)^T, \label{pow}
\end{eqnarray}
where $\mathcal{C}$ denotes the {\emph closure} operation $\mathcal{C}\boldsymbol{q}:=\boldsymbol{q}/\sum_iq_i$. An inverse perturbation is given by $\ominus\boldsymbol{q}:=\oplus(-1)\odot\boldsymbol{q}$.
}

\subsection{Power Transformed Compositions as Convex Combinations, Dual Geodesics}

The shrinkage estimator (\ref{shrink2}) is a weighted mean of the target and the observed point. This convex combination is an example for what is known as a mixture geodesic (or $m$-geodesic) in information geometry. Consider now a similar structure using the operations of perturbation and powering introduced above:
\begin{equation}
\tilde{\boldsymbol{q}}=\lambda\odot\boldsymbol{\tau}\oplus(1-\lambda)\odot\hat{\boldsymbol{q}}.\label{expogeo_formal}
\end{equation}
This describes a so-called \emph{exponential} geodesic (or $e$-geodesic).\footnote{This is also known as the \emph{Hellinger arc} connecting two distributions.} Usually \cite{Amari}, both types of geodesics are written in terms of their dual coordinates:
\begin{eqnarray}
    \boldsymbol{\eta}(\lambda)&=&\lambda\boldsymbol{\eta}_{\boldsymbol{\tau}}+(1-\lambda)\boldsymbol{\eta}_{\hat{\boldsymbol{q}}},\label{mixgeo}\\
    \boldsymbol{\theta}(\lambda)&=&\lambda\boldsymbol{\theta}_{\boldsymbol{\tau}}+(1-\lambda)\boldsymbol{\theta}_{\hat{\boldsymbol{q}}},\label{expogeo}
\end{eqnarray}
where we used subscripts to indicate at which points the coordinates are evaluated. Coming back to the power-transformation (\ref{powering}), we can easily see that it is described by the exponential geodesic between the observed point and the uniform target: Evaluating the exponential coordinates at $f_\beta(\hat{\boldsymbol{q}})$, we have 
\begin{equation}
    \boldsymbol{\theta}_{f_\beta(\hat{\boldsymbol{q}})}=\left(\log\frac{\hat{q}_1^\beta}{\hat{q}_D^\beta},\dots,\log\frac{\hat{q}_{D-1}^\beta}{\hat{q}_D^\beta}\right)^T=\beta\boldsymbol{\theta}_{\hat{\boldsymbol{q}}}.
\end{equation}
We also notice that for $\boldsymbol{\tau}=(1/D)_{i=1}^D$, $\boldsymbol{\theta}_{\boldsymbol{\tau}}$ vanishes. Setting $\beta=1-\lambda$, we immediately obtain (\ref{expogeo}). When evaluating (\ref{expogeo}) for a general target, we can use the form (\ref{expogeo_formal}) to obtain a generalized power transformation in terms of the original parameters:
\begin{equation}
   \hat{\boldsymbol{q}}^\mathrm{es} := \left(\frac{\tau_1^{1-\beta} \hat{q}_1^{\beta}}{\sum_{k=1}^{D}\tau_k^{1-\beta} \hat{q}_k^{\beta}},\dots,\frac{\tau_D^{1-\beta} \hat{q}_D^{\beta}}{\sum_{k=1}^{D}\tau_k^{1-\beta} \hat{q}_k^{\beta}}\right)^T. \label{powering_general}
\end{equation}
Comparing $\hat{\boldsymbol{q}}^\mathrm{es}$ with the shrinkage estimator (\ref{shrink1}), we see that instead of a weighted arithmetic mean between the target and the empirical estimator, here we evaluate a weighted geometric mean between them.

\subsection{Another Reparametrization of the Posterior}

Since the generalized power transformation (\ref{powering_general}) can be described as a convex combination in exponential coordinates, it shares a structural similarity with the shrinkage estimator (\ref{shrink1}), which is obtained from a convex combination of expectation (a.k.a.\ mixture) coordinates. To make this a shrinkage problem, however, we need the resulting quantity $\hat{\boldsymbol{q}}^\mathrm{es}$ to be interpreted as an estimator. Here we argue that  {$\hat{\boldsymbol{q}}^\mathrm{es}$ is simply a reparametrization  of $\hat{\boldsymbol{q}}^\mathrm{sh}$ similar to (\ref{shrink2}). There, we went from $\mathcal{C}(\boldsymbol{n}+\boldsymbol{\alpha})$ to an expression involving $\lambda$, $\boldsymbol{\tau}$, and $\hat{\boldsymbol{q}}$. We also showed a simple reparametrization of the posterior of $\boldsymbol{\theta}$ in terms of $\hat{\boldsymbol{q}}^\mathrm{sh}$ together with the posterior sample size $\hat{n}$, see (\ref{posteriorSH}). Such alternative ways of writing posterior and posterior expectation can be obtained using $\hat{\boldsymbol{q}}^\mathrm{es}$ as well, as we will show in the following.\\
As we have seen in the previous section, an alternative parameter $\beta$ can be used to define a geometric mean between target and observed point. Defining $\tilde{n}:=\sum_{k=1}^{D}\tau_k^{1-\beta} n_k^{\beta}$, in the expression for the posterior (\ref{posteriorSH}) we can simply replace $\hat{n}\hat{\boldsymbol{q}}^\mathrm{sh}$ by new Dirichlet parameters $\tilde{n}\hat{\boldsymbol{q}}^\mathrm{es}$ to obtain the following expression of the posterior:}
\begin{equation}
    p(\boldsymbol{\theta}\mid\hat{\boldsymbol{q}}^\mathrm{es},\tilde{n})=\exp\left(\tilde{n}\left[\sum_{k=1}^{D-1}\theta^k\hat{q}^\mathrm{es}_k-\psi(\boldsymbol{\theta})\right]-\log B\left(\tilde{n}\hat{\boldsymbol{q}}^\mathrm{es}\right)\right).\label{posteriorES}
\end{equation}
{This provides us with another example for Proposition \ref{KLMAP}. Maximizing the posterior thus corresponds to a minimization of the KL-divergence between $\hat{\boldsymbol{q}}^\mathrm{es}$ and the true parameter. Furthermore, the derivation of (\ref{dataprob}) given in the Appendix also shows that $B(\hat{n}\hat{\boldsymbol{q}}^\mathrm{es})$ normalizes (\ref{posteriorSH})}.\footnote{  As $\hat{n}\hat{\boldsymbol{q}}^\mathrm{sh}=\boldsymbol{n}+\boldsymbol{\alpha}$, and $\tilde{n}\hat{\boldsymbol{q}}^\mathrm{es}$ has the exact same form.} Note that this also implies that the posterior expectation of $\boldsymbol{q}$ can be written equally valid as either the shrinkage estimator $\hat{\boldsymbol{q}}^\mathrm{sh}$ or as  the \emph{exponential} shrinkage estimator $\hat{\boldsymbol{q}}^\mathrm{es}$. {This means that the exponential shrinkage estimator is nothing but the reparametrized posterior expectation of $\boldsymbol{q}$.}

\subsection{Quadratic Risk on the Tangent Space}
\begin{figure}[h]%
\centering
\includegraphics[width=1\textwidth]{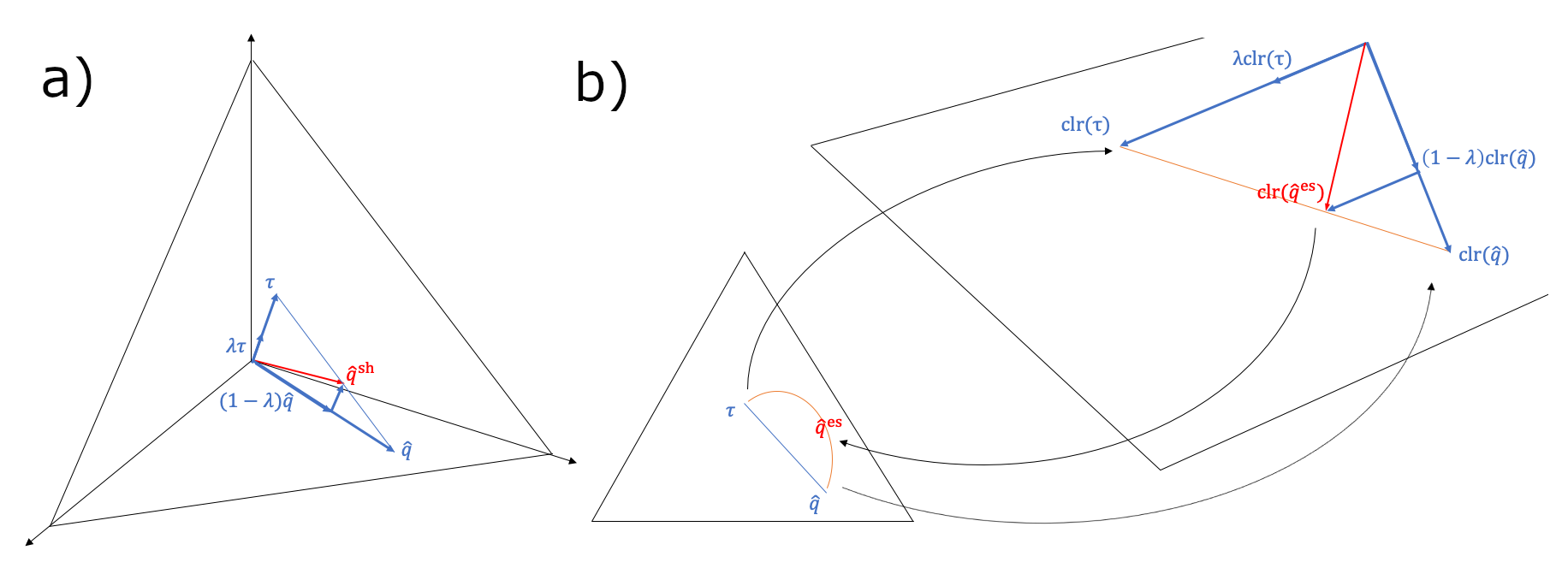}
\caption{{\bf a)} The shrinkage estimator $\hat{\boldsymbol{q}}^\mathrm{sh}$ (in red) obtained by an addition of scaled vectors (in blue) ending in the unit simplex (shown in black). The $m$-geodesic connecting $\boldsymbol{\tau}$ and $\hat{\boldsymbol{q}}$ is shown as a thin blue line. {\bf b)} The exponential shrinkage estimator $\hat{\boldsymbol{q}}^\mathrm{es}$ (in red) obtained by vector addition in the tangent space. The $e$-geodesic is shown as a curved orange line in the simplex and a straight orange line in the tangent space.}\label{shrink_simplex}
\end{figure}

To evaluate the accuracy of the exponential shrinkage estimator, we would like a simple risk function like the MSE. {We saw previously that with this risk function, an analytic estimate of the optimal prior weight was essentially possible because of the linearity of the shrinkage estimator. However, a generalized notion of linearity is now needed:} While $m$-geodesics are straight lines in the simplex, $e$-geodesics are straight lines in its tangent space
\begin{equation}
\mathcal{T}^D=\left\{\boldsymbol{v}\in\mathbb{R}^D:\sum_{i=1}^Dv_i=0\right\}.
\end{equation}
A mapping from the simplex to $\mathcal{T}^D$ (a.k.a.\ \emph{clr plane} in CoDA) is known as the clr transformation 
\begin{equation}
    \mathrm{clr}(\boldsymbol{q})=\left(\log\frac{q_1}{g(\boldsymbol{q})},\dots,\log\frac{q_D}{g(\boldsymbol{q})}\right)^T,
\end{equation}
where $g$ denotes the geometric mean 
$g(\boldsymbol{x}) = \left(\prod_{i = 1}^D q_i\right)^{1/D}$. 
{This mapping is fundamental in both information geometry and CoDA. The constraint that the clr components sum to zero means that the points on an exponential geodesic retain their normalization on the simplex.}\\
With this, a quadratic loss function in analogy to the one on the simplex can be obtained by first mapping the compositions in question to the tangent space and then using squared Euclidean distance again (see Fig. \ref{shrink_simplex}).
Let us first define the loss function on the tangent space for the empirical estimator:
\begin{equation}
    L_A(\boldsymbol{q},\hat{\boldsymbol{q}})=\sum_{j=1}^{D}\left(\mathrm{clr}_j(\hat{\boldsymbol{q}})-\mathrm{clr}_j(\boldsymbol{q})\right)^2.\label{lossclr}
\end{equation}
This is the (squared) Aitchison distance, i.e., an alternative expression of (\ref{Aitchison}). Via the mapping of the simplex to $\mathcal{T}^D$, the expression $\mathrm{clr}(\hat{\boldsymbol{q}})-\mathrm{clr}(\boldsymbol{q})$ can be interpreted as a difference vector between compositions \cite{ErbAy}. One can write this in form of a perturbation with the notation $\hat{\boldsymbol{q}}\ominus\boldsymbol{q}$, which makes the analogy with (\ref{MSE1}) even more compelling. The ``exponential" analogue to the MSE of section \ref{decision} is the risk function associated with the squared Aitchison loss, i.e.\ the expectation
\begin{equation}
    \Tilde{R}_{\boldsymbol{q}}(\hat{\boldsymbol{q}})=\mathbbm{E}L_A(\boldsymbol{q},\hat{\boldsymbol{q}})=\sum_{j=1}^{D}\left[\mathrm{var}\left(\mathrm{clr}_j(\hat{\boldsymbol{q}})\right)+\mathbbm{E}^2\left(\mathrm{clr}_j(\hat{\boldsymbol{q}})-\mathrm{clr}_j(\boldsymbol{q})\right)\right].\label{MSEclr}
\end{equation}
Unfortunately, in this case the bias term does not vanish for the empirical estimator, and we shall need an approximation to evaluate it.

\subsection{Optimization Along the Exponential Geodesic}

We can now use our modified risk function on the exponential shrinkage estimator, in analogy to (\ref{cost}), to minimize it with respect to $\lambda=1-\beta$. {Using Proposition \ref{LedoitWolf} with $f_j(\cdot)=\mathrm{clr}_j(\cdot)$, and $\lambda=1-\beta$, for the MSE of $\mathrm{clr}(\hat{\boldsymbol{q}}^\mathrm{es})$ we obtain}
\begin{multline}
    {R}_{\boldsymbol{q}}(\hat{\boldsymbol{q}}^\mathrm{es})=(1-\lambda)^2\sum_{j=1}^{D}\mathrm{var}\left(\mathrm{clr}_j(\hat{\boldsymbol{q}})\right)\\
   +\sum_{j=1}^{D}\bigg[\lambda\mathbbm{E}\left(\mathrm{clr}_j(\boldsymbol{\tau})-\mathrm{clr}_j(\hat{\boldsymbol{q}})\right)+\mathbbm{E}\mathrm{clr}_j(\hat{\boldsymbol{q}})-\mathrm{clr}_j(\boldsymbol{q})\bigg]^2.
\end{multline}
A solution for the minimum can be found at
\begin{equation}
    \lambda_\mathrm{min}=\frac{\sum_{j=1}^{D}\bigg[\mathrm{var}\left(\mathrm{clr}_j(\hat{\boldsymbol{q}})\right)-\mathbbm{E}\left(\mathrm{clr}_j(\boldsymbol{\tau})-\mathrm{clr}_j(\hat{\boldsymbol{q}})\right)\left(\mathbbm{E}\mathrm{clr}_j(\hat{\boldsymbol{q}})-\mathrm{clr}_j(\boldsymbol{q})\right)\bigg]}{\sum_{j=1}^{D}\mathbbm{E}\bigg[\left(\mathrm{clr}_j(\boldsymbol{\tau})-\mathrm{clr}_j(\hat{\boldsymbol{q}})\right)^2\bigg]}\label{lambdamin}
\end{equation}
Again, this can be evaluated in practice by replacing $\boldsymbol{q}$ by the best estimator available. To estimate the variance and the expectation terms of the clr-transformed empirical estimator, we resort to Taylor expansion. While the expressions become a bit more unwieldy compared with the ones on the $m$-geodesic, we can still evaluate them explicitly. For the mean we get
\begin{equation}
\mathbbm{E}\mathrm{clr}_j(\hat{\boldsymbol{q}})\approx E_j:=\mathrm{clr}_j(\boldsymbol{q})-\frac{1-q_j}{2q_jn}+\frac{1}{2D}\sum_{k=1}^D\frac{1-q_k}{q_kn},\label{E}
\end{equation}
and for the variance (where this approximation  is known as the Delta method) 
\begin{multline}
    \mathrm{var}\left(\mathrm{clr}_j(\hat{\boldsymbol{q}})\right)\approx\\ V_j:=\left(1-\frac{2}{D}\right)\frac{1-q_j}{q_jn}+\frac{1}{D^2}\sum_{k=1}^D\frac{1-q_k}{q_kn}-\frac{1}{n}\left(3-\frac{7}{D}+\frac{4}{D^2}\right)\label{V}
\end{multline}
(see Appendix for a derivation). In the case of the maximum-entropy target, the clr$_j(\boldsymbol{\tau})$ terms in (\ref{lambdamin})  vanish, and an estimator of the optimal power can be obtained by
\begin{equation}
    \beta^*=1-\frac{\sum_{k=1}^D\left[V_k-E_k(E_k-\mathrm{clr}_k(\boldsymbol{q}))\right]}{\sum_{k=1}^D\left[V_k+E_k^2\right]}.
\end{equation}

\subsection{Performance on Simulated Data}
We can now test how well we can infer true frequencies from simulated data using the exponential shrinkage estimator. For this, we use the equidistribution as the target and optimize the $\beta$ parameter as described before. 
\begin{figure}[h]%
\centering
\includegraphics[width=1\textwidth]{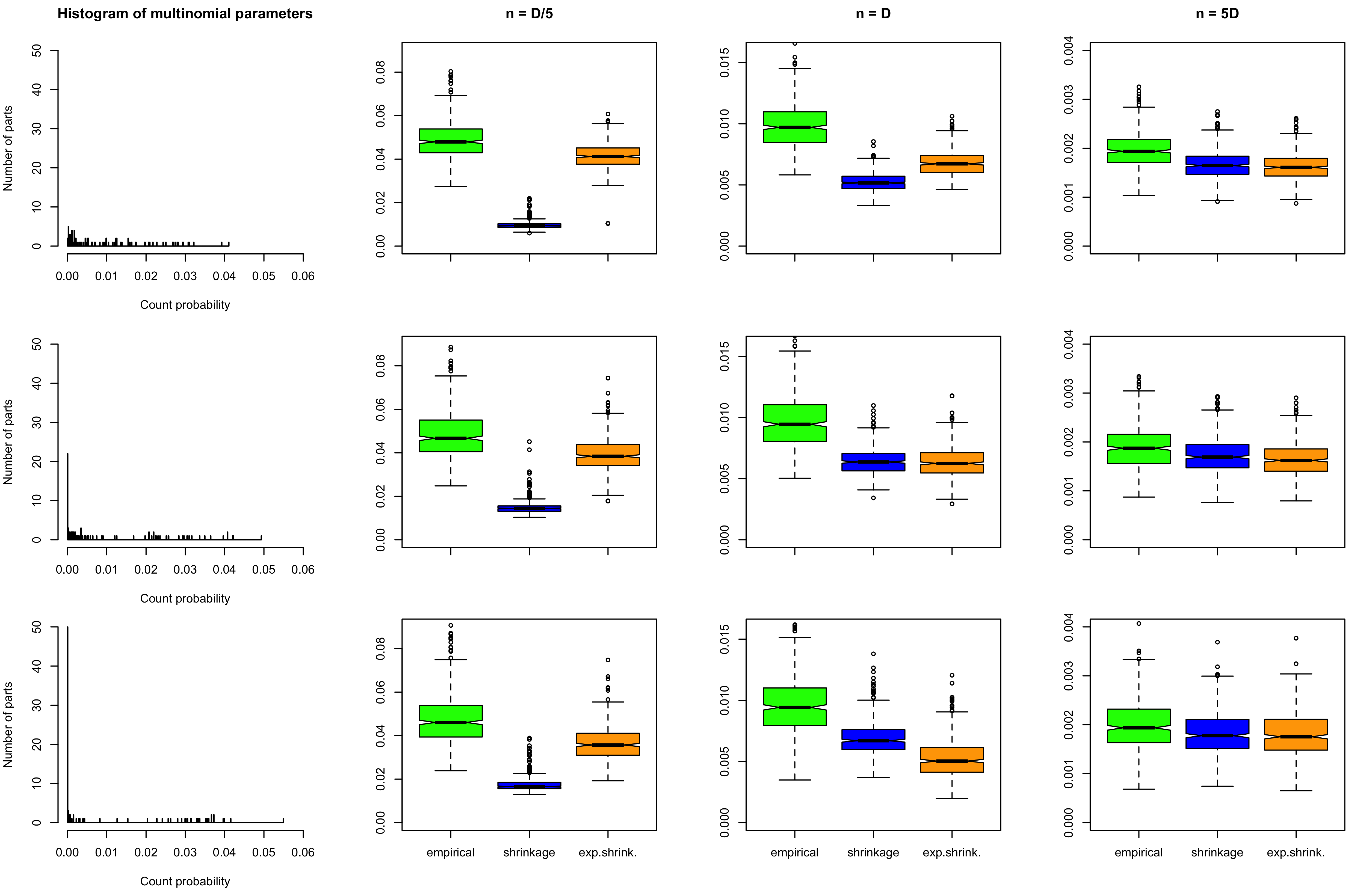}
\caption{Mean squared error (MSE) of the empirical estimator (green), the shrinkage estimator (blue), and the exponential shrinkage estimator (orange). Data are sampled from multinomial distributions with increasing sparsity. Boxplots in each row show the MSEs of 500 samples from the multinomial whose histogram is shown in the first column. Sample size increases from left to right ($n=D/5,D,5D$), while sparsity increases from top down. As $D=100$, the vertical axis in the histograms can be read as a percentage. Note that the vertical boxplot axes change their range between columns.}\label{simulation}
\end{figure}
This should not be understood as an intent at a comprehensive benchmark but rather as a proof of concept. We test performance on multinomial counts only. The three different multinomial distributions ($D$=100) shown in Figure \ref{simulation} were obtained by sampling from Dirichlet distributions with three different choices for the hyper parameters. These were chosen to obtain multinomial parameters that are far from equidistributed and have an increasing number of essential zeros. As a measure of performance, we chose MSE as in \cite{HausserStrimmer}. Beside being simple and intuitive, MSE has the advantage that zeros are not problematic as there are no logarithms involved. Both zeros as obtained from undersampling (i.e., \emph{count} zeros) as well as those that occur because parameters are truly (or almost) zero (so-called \emph{essential} zeros) will have the effect that the observed point $\hat{\boldsymbol{q}}$ falls on the boundary of the simplex. This is not a problem for the shrinkage estimator, as $m$-geodesics can go from the centre to the boundary. However, $e$-geodesics are only defined inside the simplex, and we have to redefine the observed point as its projection to the nonzero parts, with a subsequent change in the dimension $D$. In any case, it is only the nonzero parts that can be modified by the exponential shrinkage estimator. As an approximation of the true parameter in the expressions (\ref{E}) and (\ref{V}), we use the shrinkage estimator $\hat{\boldsymbol{q}}^\mathrm{sh}$. The exponential shrinkage estimator is optimized over the nonzero parts only. The results show that exponential shrinkage outperforms the empirical estimator but cannot compete with the shrinkage estimator if the data are severely undersampled (first column in Figure \ref{simulation}). There is a sweet spot of performance when many essential zeros are present and the data are sampled at reasonable depth (middle column). In this case, the exponential shrinkage estimator can outperform the shrinkage estimator. Clearly, it is ``already correct" for the unobserved values, while the shrinkage estimator imputes them. Further increasing sample size essentially equalizes the performance of all estimators (right column). Note that the presence of zeros in the multinomial parameters effectively increases the sample size as the same counts are now distributed over fewer parts. The two factors studied in Figure \ref{simulation}, sample size and sparsity, are thus not independent of each other in their effects.

\subsection{Discussion}
We have shown that power transformations of relative count data can be understood as a shrinkage problem. An analytic solution for the optimal power for given data can be obtained in a way that is analogous to what was proposed for finding an optimal flattening constant. We find the underlying information-geometric structure intriguing: Both types of geodesics between the empirical estimate and the maximum-entropy estimate give rise to their own shrinkage problem. But we think that there are also practical implications for data anlysis. In the context of compositional data visualization, power transformations have been proposed as an approximation to log-ratio transformations, which require zero imputation. Correspondence Analysis (CA), one of the best methods for visualizing two-way tables containing counts, can be made more suitable for relative count data when applying such a transformation. It then approximates log-ratio analysis (LRA), whose visualization appeals more to our Euclidean intuition but whose zero imputed data may be suboptimal {or even impossible for very sparse data sets. For side-by-side visualizations of geochemical and single-cell data using both methods, see \cite{Reappraisal}.} While CA is a visualization of the stretched out (weighted) simplex, LRA is a PCA on its tangent space (the clr plane). When using the hybrid approach of CA with power transformed counts, currently a uniform power parameter is applied to an entire data matrix that could contain rows with heterogeneous sample sizes. As we have seen, in terms of an optimal approximation to the underlying parameters in each row, this would work best if samples follow the same distribution and the sample sizes are not too different. On the other hand, we could argue that, from a modelling perspective, it would be better to find the best power for each row in the data matrix separately. While the deformation with respect to LRA would now be heterogeneous among samples, the fit with underlying population parameters would be better. The shrinkage approach is of course applicable beyond data visualization, and we think that applying it as a kind of data normalization holds some promise for very sparse data sets as occurring in microbiome analysis or single-cell genomics. Not all of these zeros are essential zeros, but many of them may be caused by truly small occurrence probabilities. If so, the commonly applied log transform with a uniform pseudocount would almost certainly be less suitable than a data-driven power transformation as proposed here. While this approach may still appear overly simplistic, given today's highly complex data acquisition protocols where effects of statistical and engineering decisions are hard to disentangle, simple approaches often perform similarly well as highly complex ones \cite{Pachter}.

\section*{Appendix}
\subsection*{Derivation of Equation\ \ref{dataprob}}
Inserting the expressions (\ref{suffstat}-\ref{betafunc}) into the general conjugate prior (\ref{conprior}), we obtain
\begin{equation}
    \pi(\boldsymbol{\theta}\mid\boldsymbol{\alpha})=\exp\left(\sum_{k=1}^{D-1}\alpha_k\theta^k-\psi(\boldsymbol{\theta})\sum_{k=1}^D\alpha_k-\log B(\boldsymbol{\alpha})\right).
\end{equation}
Together with (\ref{expo2}), the denominator in (\ref{Bayes}) becomes
\begin{equation}
    p(\boldsymbol{r}\mid\boldsymbol{\alpha})=\frac{1}{B(\boldsymbol{\alpha})}\int d\boldsymbol{\theta}\exp\left(\sum_{k=1}^{D-1}\theta^k(n_k(\boldsymbol{r})+\alpha_k)-\left(n+\sum_{k=1}^D\alpha_k\right)\psi(\boldsymbol{\theta})\right).
\end{equation}
Now a variable transformation to the original parameter $\boldsymbol{q}$ with Jacobian det$(\partial\theta_j/\partial q_j)_{j=1}^{D-1}=\prod_{k=1}^Dq_k^{-1}$ gives for the integral
\begin{multline}
    B(\boldsymbol{\alpha})p(\boldsymbol{r}\mid\boldsymbol{\alpha})\\
    =\int\frac{d\boldsymbol{q}}{\prod_{k=1}^Dq_k}\exp\left(\sum_{k=1}^{D-1}(n_k+\alpha_k)\log\frac{q_k}{q_D}+\left(n+\sum_{k=1}^D\alpha_k\right)\log q_D\right)\\
    =\int\frac{d\boldsymbol{q}}{\prod_{k=1}^Dq_k}\exp\left(\sum_{k=1}^{D-1}(n_k+\alpha_k)\log q_k+(n_D+\alpha_D)\log q_D\right)\\
    =\int d\boldsymbol{q}{\prod_{k=1}^Dq_k^{n_k+\alpha_k-1}}=B\left(\boldsymbol{n}+\boldsymbol{\alpha}\right),
\end{multline}
by definition of the multivariate beta function. We shortened $n_k(\boldsymbol{r})$ to $n_k$ here.

\subsection*{Proof of Proposition \ref{KLMAP}}
By definition of $f$ we have
\begin{equation}
-\log f(\tilde{\boldsymbol{q}},\tilde{n},\boldsymbol{\theta})=-\log Z(\tilde{n},\tilde{\boldsymbol{q}})-\left\{\tilde{n}\left(\boldsymbol{\theta}\tilde{\boldsymbol{\eta}}-\psi(\boldsymbol{\theta})\right)\right\}.
\end{equation}
Using (\ref{KL}), the negative curly brackets can be replaced by $\tilde{n}(D_\phi-\phi)$, so we obtain
\begin{equation}
-\log f(\tilde{\boldsymbol{q}},\tilde{n},\boldsymbol{\theta})=-\log Z(\tilde{n},\tilde{\boldsymbol{q}})+\tilde{n}\left(D(\tilde{\boldsymbol{q}}\mid\mid\boldsymbol{q})-\phi(\tilde{\boldsymbol{\eta}})\right). 
\end{equation}
Rearranging terms, we obtain the proposition:
\begin{equation}
\tilde{n}D(\tilde{\boldsymbol{q}}\mid\mid\boldsymbol{q})=\tilde{n}\phi(\tilde{\boldsymbol{\eta}})+\log Z(\tilde{n},\tilde{\boldsymbol{q}})-\log f(\tilde{\boldsymbol{q}},\tilde{n},\boldsymbol{\theta}).
\end{equation}

\subsection*{Proof of Proposition \ref{LedoitWolf}}
(i) Using the bias-variance decomposition as in (\ref{MSE1}), for the quadratic risk of $\tilde{\boldsymbol{q}}$ we obtain
\begin{multline}
    R_{\boldsymbol{q}}(\tilde{\boldsymbol{q}})=\mathbbm{E}(\tilde{\boldsymbol{q}}-f(\boldsymbol{q}))^2=\\
    \sum_{j=1}^{D}\mathrm{var}\big(\lambda f_j(\boldsymbol{\tau})+(1-\lambda)f_j(\hat{\boldsymbol{q}})-f_j(\boldsymbol{q})\big)+\sum_{j=1}^{D}\mathbbm{E}^2\big(\lambda f_j(\boldsymbol{\tau})+(1-\lambda)f_j(\hat{\boldsymbol{q}})-f_j(\boldsymbol{q})\big)\\
    =\sum_{j=1}^{D}\left[\lambda^2\mathrm{var}\big(f_j(\boldsymbol{\tau})\big)+(1-\lambda)^2\mathrm{var}\big(f_j(\hat{\boldsymbol{q}})\big)+2\lambda(1-\lambda)\mathrm{cov}\big(f_j(\boldsymbol{\tau}),f_j(\hat{\boldsymbol{q}}\big)\right]\\
    +\sum_{j=1}^{D}\mathbbm{E}^2\left(\lambda(f_j(\boldsymbol{\tau})-f_j(\hat{\boldsymbol{q}}))+f_j(\hat{\boldsymbol{q}})-f_j(\boldsymbol{q})\right)\\
    =(1-\lambda)^2\sum_{j=1}^{D}\mathrm{var}\big(f_j(\hat{\boldsymbol{q}})\big)+\sum_{j=1}^{D}\bigg(\mathbb{E}f_j(\hat{\boldsymbol{q}})-f_j(\boldsymbol{q})-\lambda\big(\mathbb{E}f_j(\hat{\boldsymbol{q}})-f_j(\boldsymbol{\tau})\big)\bigg)^2.
\end{multline}
For the first equality, the variance of the sum is evaluated in the usual way as a quadratic form. We can ignore the $f_j(\boldsymbol{q})$ term because it is constant. Similarly, the last equality uses the fact that the $f_j(\boldsymbol{\tau})$ are fixed parameters, so their variance and covariance terms vanish, showing the first part of the proposition.\\
(ii) To obtain the minimum of the cost function, we derive by $\lambda$ and set the result zero (while the second derivative is always greater 0):
\begin{multline}
    \frac{dR_{\boldsymbol{q}}(\tilde{\boldsymbol{q}})}{d\lambda}=-2(1-\lambda)\sum_{j=1}^{D}\mathrm{var}\big(f_j(\hat{\boldsymbol{q}})\big)\\
    -2\sum_{j=1}^{D}\big(\mathbb{E}f_j(\hat{\boldsymbol{q}})-f_j(\boldsymbol{\tau})\big)\bigg(\mathbb{E}f_j(\hat{\boldsymbol{q}})-f_j(\boldsymbol{q})-\lambda\big(\mathbb{E}f_j(\hat{\boldsymbol{q}})-f_j(\boldsymbol{\tau})\big)\bigg)=0.
\end{multline}
From this it follows that
\begin{multline}
   \sum_{j=1}^{D}\mathrm{var}\big(f_j(\hat{\boldsymbol{q}})\big)+\sum_{j=1}^{D}\big(\mathbb{E}f_j(\hat{\boldsymbol{q}})-f_j(\boldsymbol{\tau})\big)\big(\mathbb{E}f_j(\hat{\boldsymbol{q}})-f_j(\boldsymbol{q})\big)\\
   =\lambda\sum_{j=1}^{D}\mathrm{var}\big(f_j(\hat{\boldsymbol{q}})\big)+\lambda\sum_{j=1}^{D}\big(\mathbb{E}f_j(\hat{\boldsymbol{q}})-f_j(\boldsymbol{\tau})\big)^2.
\end{multline}
Finally, using the fact that $\mathrm{var}\big(f_j(\hat{\boldsymbol{q}})\big)+\big(\mathbb{E}f_j(\hat{\boldsymbol{q}})-f_j(\boldsymbol{\tau})\big)^2$= $\mathbbm{E}\left[\big(f_j(\hat{\boldsymbol{q}})-f_j(\boldsymbol{\tau})\big)^2\right]$, we obtain (ii), concluding the proof.

\subsection*{Expectation and Variance of the CLR-Transformed Empirical Estimator}

Consider the Taylor expansion of the $j$-th component of clr($\hat{\boldsymbol{q}}$) around $\boldsymbol{q}$ up to second order terms
\begin{multline}
    \mathrm{clr}_j(\hat{\boldsymbol{q}})\approx\mathrm{clr}_j(\boldsymbol{q})\\
    +\sum_{k=1}^D\frac{\partial\mathrm{clr}_j(\boldsymbol{q})}{\partial q_k}(\hat{q}_k-q_k)+\frac{1}{2}\sum_{k,l}\frac{\partial^2\mathrm{clr}_j(\boldsymbol{q})}{\partial q_k\partial q_l}(\hat{q}_k-q_k)(\hat{q}_l-q_l).\label{Taylor}
\end{multline}
The first derivatives evaluate to
\begin{equation}
      \frac{\partial\mathrm{clr}_j(\boldsymbol{q})}{\partial q_k}=
    \left\{
      \begin{array}{c@{\quad}l}
         \frac{1-1/D}{q_j} & \mbox{if $j=k$,} \\
          \frac{1}{Dq_k} 
         & \mbox{if $j\ne k$,}
      \end{array}
    \right. \label{1st}
\end{equation}
and the second derivatives are 
\begin{equation}
      \frac{\partial^2\mathrm{clr}_j(\boldsymbol{q})}{\partial q_k\partial q_l}=
    \left\{
      \begin{array}{c@{\quad}l}
         -\frac{1-1/D}{q_j^2} & \mbox{if $j=k=l$,} \\
          \frac{1}{Dq_k^2} 
         & \mbox{if $j\ne k=l$,}\\
         0 & \mbox{else.}
      \end{array}
    \right. 
\end{equation}
When taking the expectation of (\ref{Taylor}), the first-order terms vanish due to the linearity of expectation. In the second-order terms, only those where $k=l$ remain. We thus obtain 
\begin{equation}
    \mathbbm{E}\mathrm{clr}_j(\hat{\boldsymbol{q}})\approx\mathrm{clr}_j(\boldsymbol{q})-\frac{\mathbbm{E}(\hat{q}_j-q_j)^2}{2q_j^2}+\frac{1}{2D}\sum_{k=1}^D\frac{\mathbbm{E}(\hat{q}_k-q_k)^2}{q_k^2}.\label{E2}
\end{equation}
Now using the bias-variance decomposition (\ref{MSE1}), we have
\begin{equation}
    \mathbbm{E}(\hat{q}_j-q_j)^2=\mathrm{var}(\hat{q}_j)=\frac{1}{n^2}\mathrm{var}(n_j)=\frac{q_j(1-q_j)}{n}.\label{var}
\end{equation}
Inserting this into (\ref{E2}), we obtain (\ref{E}). Similarly, for the variance of the clr-transformed empirical estimator, we evaluate the variance of (\ref{Taylor}). The 0-th order does not contribute because it is non-stochastic, and we ignore the second-order terms as commonly done using the Delta method. The variance $V_j$ of the first order terms evaluates to
\begin{equation}
    V_j=\mathrm{var}\left(\sum_{k=1}^D\frac{\partial\mathrm{clr}_j(\boldsymbol{q})}{\partial q_k}\hat{q}_k\right)= \sum_{k,l}\left(\frac{\partial\mathrm{clr}_j(\boldsymbol{q})}{\partial q_k}\right)\left(\frac{\partial\mathrm{clr}_j(\boldsymbol{q})}{\partial q_l}\right)\mathrm{cov}(\hat{q}_k,\hat{q}_l),\label{varclr}
\end{equation}
by evaluating the square and using bilinearity of covariance. The covariance elements for equal indices are given in (\ref{var}). The off-diagonal terms are
\begin{equation}
    \mathrm{cov}(\hat{q}_k,\hat{q}_l)=\mathrm{cov}\left(\frac{n_k}{n},\frac{n_l}{n}\right)=\frac{1}{n^2}\mathrm{cov}(n_k,n_l)\stackrel{k\ne l}{=}\frac{-q_kq_l}{n},
\end{equation}
by the well-known expression in the multinomial case. We now collect the respective covariance terms and first derivatives to evaluate (\ref{varclr}). The double sum decomposes into four terms that correspond to the cases where the indices are not equal and don't contain $j$, are equal and don't contain $j$, are not equal and one of them is $j$, and are both equal to $j$, respectively:
\begin{multline}
   V_j=\sum_{k\ne j}\sum_{l\ne j,\atop l\ne k}\frac{-q_kq_l}{D^2q_kq_ln}\\
    +\sum_{l\ne j}\frac{q_l(1-q_l)}{D^2q_l^2n} +
    2\sum_{k\ne j}\frac{(1-1/D)(-q_kq_j)}{Dq_kq_jn}+\frac{(1-1/D)^2q_j(1-q_j)}{q_j^2n}\\
    =\frac{-(D-1)(D-2)}{D^2n}+\sum_{l\ne j}\frac{1-q_l}{D^2q_ln}-2\frac{(1-1/D)(D-1)}{Dn}\\
    +\frac{(1-1/D)^2(1-q_j)}{q_jn}.
\end{multline}
This can be further simplified including a part of the last term in the summation of the second term (getting rid of $l\ne j$) and joining the two terms independent of $\boldsymbol{q}$ into a single expression. After this we obtain
\begin{multline}
    V_j=(1-2/D)\frac{1-q_j}{q_jn}+\frac{1}{D^2}\sum_{l=1}^D\frac{1-q_l}{q_ln}\\
    -\frac{1}{n}\left(\frac{(D-1)(D-2)}{D^2}+2\frac{(1-1/D)(D-1)}{D}\right).
\end{multline}
With further simplification of the last term, this is (\ref{V}).

\section*{Declarations}

{\bf Data availability:} R scripts for synthetic data and analysis can be downloaded from the GitHub repository ionase/exshrink.\\
{\bf Conflict of interest:} I declare that there is no conflict of interest.\\
{\bf Acknowledgements:} I thank Nihat Ay for helpful comments on an early version of the manuscript. An anonymous reviewer's suggestions led to further improvements.

\end{document}